\documentclass[aps,twocolumn,10pt,prl]{revtex4-1}
\usepackage{amsfonts}
\usepackage{amsmath}
\usepackage{amssymb}
\usepackage{amsthm}
\usepackage{array}
\usepackage{epsfig}
\usepackage{graphicx}
\usepackage{bm}%
\usepackage{verbatim}
\usepackage{pstricks,pst-node}
\usepackage{tensor}
\usepackage{tikz}
\usepackage{longtable}
\DeclareGraphicsExtensions{.eps,.ps,.pdf}
\theoremstyle{plain}
\newtheorem{theorem}{Theorem}

\theoremstyle{definition}

\theoremstyle{remark}

\newcommand{\ket}[1]{| #1 \rangle}
\newcommand{\bra}[1]{\langle #1 |}
\newcommand{\braket}[2]{\langle #1 | #2 \rangle}
\newcommand{\pket}[1]{[ #1 ]}

\newcommand{\Tr}{\text{Tr}}

\newcommand{\eg}{\hbox{\em e.g.{}}}

\newcommand{\ie}{\hbox{\em i.e.{}}}

\newcommand{\rhs}{\hbox{r.h.s.{}}}

\newcommand{\Ps}{\mathbb{P}}
\newcommand{\Hs}{\mathcal{H}}

\makeatletter
\g@addto@macro\bfseries{\boldmath}
\makeatother

\begin{document}

\title{When geometric phases turn topological}

\author{P.{} Aguilar}
\email{pedro.aguilar@nucleares.unam.mx}	
\affiliation{Instituto de Ciencias Nucleares \\
	Universidad Nacional Aut\'onoma de M\'exico\\
	PO Box 70-543, 04510, CDMX, M\'exico.}

\author{C.{} Chryssomalakos}
\email{chryss@nucleares.unam.mx}
\affiliation{Instituto de Ciencias Nucleares \\
	Universidad Nacional Aut\'onoma de M\'exico\\
	PO Box 70-543, 04510, CDMX, M\'exico.}

\author{E.{} Guzm\'an-Gonz\'alez}
\email{edgar.guzman@correo.nucleares.unam.mx}
\affiliation{Instituto de Ciencias Nucleares \\
	Universidad Nacional Aut\'onoma de M\'exico\\
	PO Box 70-543, 04510, CDMX, M\'exico.}

\author{L.{} Hanotel}
\email{hanotel@correo.nucleares.unam.mx}	
\affiliation{Instituto de Ciencias Nucleares \\
	Universidad Nacional Aut\'onoma de M\'exico\\
	PO Box 70-543, 04510, CDMX, M\'exico.}

\author{E.{} Serrano-Ens\'astiga}
\email{eduardo.serrano-ensastiga@uni-tuebingen.de}
\affiliation{Institut f\"ur Theoretische Physik \\
	Universit\"at T\"ubingen\\
	72076, T\"ubingen, Germany}

\begin{abstract}
	\noindent 
Geometric phases, accumulated when a quantum system traces a cycle in quantum state space,  do not depend on the parametrization of the cyclic  path, but do depend on the path itself. In the presence of noise that deforms the path, the phase gets affected,  compromising the robustness of possible applications, \emph{e.g.}, in quantum computing. We show that for a special class of spin states, called anticoherent, and for paths that correspond to a sequence of rotations in physical space, 
the phase only depends on topological characteristics of the path, in particular, its homotopy class, and is therefore immune to noise. 
\end{abstract}

\maketitle

%\tableofcontents

%%%%%%%%%%%%%%%%%%%%%%%%%%%%%%%%%%%%%%%%%%%%%%%%%%%%%%%%%%%
%%%%%%%%%%%%%%%%%%%%%%%%%%%%%%%%%%%%%%%%%%%%%%%%%%%%%%%%%%%
\section{Introduction}
\label{Intro}
%%%%%%%%%%%%%%%%%%%%%%%%%%%%%%%%%%%%%%%%%%%%%%%%%%%%%%%%%%%
%%%%%%%%%%%%%%%%%%%%%%%%%%%%%%%%%%%%%%%%%%%%%%%%%%%%%%%%%%%

Cyclic evolution of a quantum system gives rise to a geometric phase $\varphi_\text{geo}$, invariant under time-reparametrizations of the path $C(t)$ taken by the state of the system in the corresponding projective Hilbert space $\Ps(\Hs) \equiv \Ps$~\cite{Ber:84,Sim:83,Aha.Ana:87,Ber:89,Sam.Bha:88,Muk.Sim:93}. Motivated by this, proposals have been made to use such phases in quantum computation, the invariance mentioned translating, in this case, in noise resilience (see, \eg,~\cite{Zan.Ras:99,Jon.Ved.Eke.Cas:00,Pac.Zan:01,Ota.Kon:09}). However, whatever noise exists, affects not only the time-parametrization of $C(t)$, but also its form, and this residual effect will, in general, leave its imprint on $\varphi_\text{geo}$, to the detriment of the robustness of the computation. As a concrete mental picture, consider a spin-1/2 particle in the presence of a time-varying magnetic field $\mathbf{B}(t)=\mathbf{B}_0(t)=B \hat{\mathbf{n}}_0(t)$, starting out, at $t=0$, in the eigenstate $\ket{\hat{\mathbf{n}}_0(0) +}$ and following adiabatically $\mathbf{B}_0(t)$ for $t>0$. For a cyclic evolution, $\hat{\mathbf{n}}_0(T)=\hat{\mathbf{n}}_0(0)$, the phase $\varphi_\text{geo}$ is proportional to the area enclosed by the curve  $C_0(t)=\hat{\mathbf{n}}_0(t)$ on the unit sphere. In the presence of noise in the field, $\mathbf{B}(t)=\mathbf{B}_0(t)+\epsilon \mathbf{B}_1(t)$, the component of $\mathbf{B}_1$ along the tangent to $C_0(t)$ induces reparametrization of $C_0(t)$ and, hence, leaves $\varphi_\text{geo}$ invariant, but the normal component changes the shape of $C_0(t)$ and modifies, in general, $\varphi_\text{geo}$. The statistics of  $\varphi_\text{geo}$ were first computed analytically in~\cite{DeC.Pal:03}, considering $\mathbf{B}_1$ as a classical stochastic field, with subsequent numerical~\cite{Fil:08} and experimental~\cite{Fil.etal:09} confirmations of the results, while the subtler case of $\mathbf{B}_1$ being a quantum operator was treated in~\cite{Agu.Chr.Guz:16}, with some qualitative differences showing up (see also~\cite{Fue.Car.Bos.Ved:02} for earlier work treating $\mathbf{B}$ as a quantum field, \cite{Liu.Fen.Wan:11,Lar:12,Wan.Wei.Lia:15} for subsequent theoretical debate, and the recent experimental observation reported in~\cite{Gas.Ber.Abd.Pec.Fil.Wal:16}). 

Our focal point in this letter is to outline a scenario in which the presence of parametric noise has no effect whatsoever on the geometric phase. The mechanism presented, and possible generalizations to the non-abelian case~\cite{Wil.Zee:84,Ana:88}, could be of interest in holonomic quantum computing~\cite{Zan.Ras:99}. The setup is best presented via Majorana's stellar representation and involves anticoherent spin states --- both concepts are now briefly explained. 
%%%%%%%%%%%%%%%%%%%%%%%%%%%%%%%%%%%%%%%%%%%%%%%%%%%%%%%%%%%
%%%%%%%%%%%%%%%%%%%%%%%%%%%%%%%%%%%%%%%%%%%%%%%%%%%%%%%%%%%
%\section{Majorana costellations and anticoherent spin states}
%\label{Mcaass}
%%%%%%%%%%%%%%%%%%%%%%%%%%%%%%%%%%%%%%%%%%%%%%%%%%%%%%%%%%%
%%%%%%%%%%%%%%%%%%%%%%%%%%%%%%%%%%%%%%%%%%%%%%%%%%%%%%%%%%%
In a little known 1932 paper~\cite{Maj:32}, Majorana pointed out that spin-$s$ quantum states can be labeled by a constellation of $2s$ points on the unit sphere. The recipe given is simple as it is cryptic: expand the state in question $\ket{\psi}$ in the $S_z$ eigenbasis, $\ket{\psi}=\sum_{k=-s}^s c_k \ket{s,k}$, and use the expansion coefficients to write down a polynomial in an auxiliary variable $\zeta$,
\begin{equation}
\label{Ppsidef}
P_{\psi}(\zeta)
=
\sum_{k=0}^{2s} (-1)^{2s-k} c_k \sqrt{\binom{2s}{k}} \zeta^k
\, ,
\end{equation}
the roots of which, stereographically projected from the south pole onto the unit sphere $S^2$, supply the \emph{Majorana constellation} of $\ket{\psi}$. When $\ket{\psi}$ is transformed in Hilbert space by the unitary representation $D(R)$ of a rotation  $R \in SO(3)$, the constellation rotates by $R$ on $S^2$. Additional information about the Majorana constellation may be found in~\cite{Ben.Zyc:17,Chr.Guz.Ser:18} while interesting applications appear in, \eg,~\cite{Bar.Tur.Dem:06,Bar.Tur.Dem:07,Mak.Suo:07}.

A \emph{spin coherent} state $\ket{\mathbf{n}}$, where $\mathbf{n}$ lives on the 2-sphere $S^2$, is an eigenstate of $\mathbf{n} \cdot \mathbf{S}$ with eigenvalue $s$ ($\mathbf{S}$ denotes the spin-$s$ representation of the generators of $\mathfrak{su}(2)$)~\cite{Per:86,Rad:71,Chr.Guz.Ser:18}. Spin coherent states maximize the modulus of the spin expectation value and are, in many respects, the ``most classical'' spin states available. Their Majorana constellation consists in $2s$ coincident stars in the direction $\mathbf{n}$ --- intuitively, they are the most directional states possible. In~\cite{Zim:06}, Zimba considered the natural question of which spin states should be declared the ``most quantum'', or, as he aptly named them, ``anticoherent'' --- one would expect these to correspond to constellations spread out ``as uniformly as possible'' over $S^2$. Analytically, the natural requirement, adopted by Zimba, for an anticoherent state $\ket{\psi}$ is that its spin expectation value vanish, $\bra{\psi} \mathbf{n} \cdot \mathbf{S} \ket{\psi}=0$, for all $\mathbf{n}$ in $S^2$. At around the same time, those same states showed up in the classification of the different phases of bosonic condensates~\cite{Bar.Tur.Dem:06,Bar.Tur.Dem:07}, as well as the study of the so called inert states~\cite{Mak.Suo:07}.
%%%%%%%%%%%%%%%%%%%%%%%%%%%%%%%%%%%%%%%%%%%%%%%%%%%%%%%%%%%
%%%%%%%%%%%%%%%%%%%%%%%%%%%%%%%%%%%%%%%%%%%%%%%%%%%%%%%%%%%
\section{When geometric phases turn topological}
\label{Wgptt}
%%%%%%%%%%%%%%%%%%%%%%%%%%%%%%%%%%%%%%%%%%%%%%%%%%%%%%%%%%%
%%%%%%%%%%%%%%%%%%%%%%%%%%%%%%%%%%%%%%%%%%%%%%%%%%%%%%%%%%%
The formulation of geometric phases we adopt here is the one put forth in~\cite{Muk.Sim:93}, which generalizes the original setup of~\cite{Ber:84} to the case of non-cyclic, non-adiabatic evolutions. Given a curve $C(t)=[\psi(t)]$  in $\mathbb{P}$, with $0 \leq t \leq 1$, a geometric phase $\varphi_\text{geo}$ may be associated to it via
\begin{align}
\label{gpMS}
\varphi_\text{geo}
&=
\varphi_\text{tot}+\varphi_{\text{dyn}}
\nonumber
\\
&=
\arg \braket{\psi(0)}{\psi(1)}
+
i\int_\mathcal{C} dt \, \bra{\psi(t)} \partial_t \ket{\psi(t)}
\, ,
\end{align}
where $\mathcal{C}(t) \equiv \ket{\psi(t)}$ is an arbitrary lift of $C(t)$ in the Hilbert space $\mathcal{H}$, and the two terms  on the right hand side are known as the \emph{total} and \emph{dynamical} phase, respectively. It can be shown easily  that $\varphi_\text{geo}$ does not depend on the lift, and is, therefore, a property of $C(t)$. It is also easy to show that $\varphi_\text{geo}$ does not change under a reparametrization of $C(t)$, $C(t) \rightarrow C'(t)=C(s(t))$, with $s(t)$ a monotonically increasing function of $t$. Under  more general perturbations of $C(t)$ though, that affect the locus of points the curve passes through, $\varphi_\text{geo}$ does change --- when such perturbations are infinitesimal, and leave the endpoints fixed, only the second term of~(\ref{gpMS}), \ie, the dynamical phase,  contributes to the change of $\varphi_\text{geo}$.

We specify now the above to the case in which $C(t)$ is the path taken by a spin-$s$ state under a sequence of rotations $R(t) \in SO(3)$, $C(t)=[\psi(t)]$, with 
\begin{equation}
\label{psitRpsi0}
\ket{\psi(t)}=D(R(t)) \ket{\psi(0)}
\, , 
\end{equation}
where $D(R(t))$ is the spin-$s$ irreducible representation of $R(t)$. Taking into account that for $D(R(t))=e^{-i \mathbf{m}(t) \cdot \mathbf{S}}$,
\begin{equation}
\label{derDRt}
\partial_t D(R(t))
=
-i (\tilde{\mathbf{m}}(t)\cdot  \mathbf{S}) \, D(R(t))
\end{equation}
holds, with $\tilde{\mathbf{m}}(t)$ a complicated function of $\mathbf{m}(t)$, the details of which are not essential to our argument, $\varphi_\text{dyn}$ in the \rhs{} of~(\ref{gpMS}) becomes
\begin{equation}
\varphi_\text{dyn}
=
\int_{t_\text{i}}^{t_\text{f}} dt \, \bra{\psi(t)}  \tilde{\mathbf{m}}(t) \cdot \mathbf{S} \ket{\psi(t)}
=
0
\label{phidynzero}
\, ,
\end{equation}
the second equality only being valid when $\ket{\psi(0)}$, and, hence, $\ket{\psi(t)}$, are anticoherent. Then $\varphi_\text{geo}$ in~(\ref{gpMS}) reduces to $\varphi_\text{tot}$, depending only on the endpoints $\ket{\psi(0)}$, $\ket{\psi(1)}$ of the lift $\mathcal{C}(t)$. It is important to keep in mind that $\mathcal{C}(t)$ is not an arbitrary lift of $C(t)$, rather, it is uniquely determined by~(\ref{psitRpsi0}) once the phase of, say, $\ket{\psi(0)}$ has been chosen. It is also clear that $\varphi_\text{geo}$ is independent of this latter choice, as the phase change $\ket{\psi(0)}\rightarrow e^{i \alpha} \ket{\psi(0)}$ implies $\ket{\psi(1)}\rightarrow e^{i \alpha} \ket{\psi(1)}$, and $\arg \braket{\psi(0)}{\psi(1)}$ remains invariant.

We consider now anticoherent states with Majorana representations that have a non-trivial rotation symmetry group $\Gamma$, which we assume to be a discrete subgroup of $SO(3)$. This latter assumption only simplifies the presentation --- our results below easily extend to the  special case in which all stars lie on a diameter of $S^2$, so that the symmetry group has a continuous $U(1)$ component. In the presence of such discrete symmetries, there are open curves $R(t)$ in $SO(3)$ that give rise, via~(\ref{psitRpsi0}), to closed curves $[\psi(t)]$ in $\mathbb{P}$. Take, for example, a curve $R(t)$ that starts, at $t=0$, at the identity $e$ of $SO(3)$, and ends, at $t=1$, at the rotation $R_m \in \Gamma$.
Since the Majorana constellation determines the state up to phase, and $R_m$ is a symmetry of the constellation, we have  (putting $\ket{\psi(0)} \equiv \ket{\Psi}$),
\begin{equation}
\label{Rspsi}
D(R_m) \ket{\Psi}=e^{i \alpha_m} \ket{\Psi}
\, ,
\end{equation}
so that $\pket{\psi(1)}=\pket{\psi(0)} =\pket{\Psi}$, \ie, $\pket{\psi(t)}$ is a closed curve in $\mathbb{P}$. For such a curve, (\ref{gpMS}), (\ref{phidynzero}) imply that $\varphi_\text{geo}=\alpha_m$, \emph{i.e.}, \emph{regardless of the  details of the curve $R(t)$, the geometrical phase only depends on its endpoints.}

In fact, $\pket{\psi(t)}$ lies in the subset of $\mathbb{P}$ given by the orbit $\mathcal{O}_{\pket{\Psi}}$ of $\pket{\Psi}$ under the action of $SO(3)$, so that $C(t)$ is a closed curve in $\mathcal{O}_{\pket{\Psi}} \subset \mathbb{P}$. The $SO(3)$ orbit of a state consists of all states that share the same shape of their Majorana constellations, but differ in its orientation in space. For a general state $\pket{\Psi}$, this orbit is a copy of $SO(3)$, but in the presence of symmetries, codified by $\Gamma$, it reduces to the quotient space $SO(3)/\Gamma$, in which two rotations $R$, $R'$ are identified if there exists a symmetry rotation $R_m \in \Gamma$ such that $R'=R R_m$. One may visualize $\mathcal{O}_{\pket{\psi}}$ as a certain subset of $SO(3)$ in an infinite number of ways --- a canonical choice is to define a biinvariant distance function $D(g_1,g_2)$ in $SO(3)$, given by $D(g_1,g_2)=\Tr (g_1 g_2^{-1})$ and then assign to each symmetry rotation $R_m \in \Gamma$ a ``cell'' $C_m$ consisting of all group elements in $SO(3)$ for which the closest symmetry rotation is $R_m$, \ie, 
\begin{equation}
\label{celldef}
\min_i(D(g,R_i))=D(g,R_m) \Rightarrow g \in C_m
\, .
\end{equation} 
This assignment divides the whole $SO(3)$ in cells, excluding those group elements that lie on the interface between two (or more) cells, \ie, are equidistant from two (or more) symmetry rotations. These latter group elements can also be ``distributed'' among a cell and its neighbors in some canonical way --- the orbit $\mathcal{O}_{\pket{\psi}}$ may then be identified with any of the cells $C_m$. 

The identifications among points of $SO(3)$ mentioned above, that give rise to $\mathcal{O}_{\pket{\Psi}}$, endow the latter  with a complicated topology, a signature feature of which is that for two given closed curves $C_1(t)$, $C_2(t)$ in $\mathcal{O}_{\pket{\Psi}}$, both of which start, at $t=0$,  and end, at $t=1$, at the same point, there may not exist a continuous map (\emph{homotopy}) that brings one into the other, fixing all along the endpoints. One then says that the two curves belong to different \emph{homotopy classes}, denoted by $[C_1]$, $[C_2]$, respectively,  and the set of all such classes forms the \emph{fundamental group} $\pi_1(\mathcal{O}_{\pket{\Psi}})$ of $\mathcal{O}_{\pket{\Psi}}$, in which the group multiplication is given by concatenation of representative curves, \ie, $[C_1] \cdot [C_2]=[C_1 \cdot C_2]$, where $(C_1 \cdot C_2)(t)$  is the curve that first goes through $C_1$ (for $0 \leq t \leq 1/2$), and then through $C_2$ (for $1/2 < t \leq 1$).  It follows from our discussion above that 
\emph{the geometric phase acquired by an anticoherent spin state $\pket{\psi(t)}$, going around a curve $C$ in $SO(3)$ that projects to a loop $\tilde{C}$ in its orbit space $\mathcal{O}_{\pket{\psi}}$, is constant on the homotopy class of $\tilde{C}$}.

The question that naturally arises now is how many homotopy classes are there, for a given discrete symmetry group $\Gamma$, and how they combine among themselves, in other words, what is the structure of the fundamental group $\pi_1(SO(3)/\Gamma)$? The following theorem, the proof of which may be found in~\cite{Mer:79}, addresses just that:
\begin{theorem}
\label{merminthm}
Let $G$ be a connected, simply connected continuous group. Let $H$ be any subgroup of $G$. Let $H_0$ be the set of points in $H$ that are connected to the identity by continuous paths lying entirely in $H$. Then $H_0$ is a normal subgroup of $H$, and the quotient group $H/H_0$ is isomorphic to the fundamental group $\pi_1(G/H)$ of the coset space $G/H$.
\end{theorem}
Since $SO(3)$ is not simply connected, as assumed of the group $G$ in the theorem, we have to pass to its universal cover $SU(2)$. To each $SO(3)$ rotation matrix there correspond two $SU(2)$ matrices, that only differ in an overall sign. To each continuous path $R(t)$ in $SO(3)$, that starts at the identity, there corresponds a unique path $\mathcal{R}(t)$ in $SU(2)$, that also starts at the identity (there is of course a second path, that starts at minus the identity). Finally, the symmetry group $\Gamma$ gets lifted to $\Gamma^C$ in $SU(2)$, which has twice as many elements. Applying now theorem~\ref{merminthm} we conclude that $\pi_1(SU(2)/\Gamma^C) \sim \Gamma^C$, since $H_0$ in the theorem contains just the identity in our case. The result makes perfect sense intuitively: one expects that homotopy classes correspond somehow to curves that start at the identity in $SO(3)$ and get to any of the symmetry rotations $R_m$ in $\Gamma$. However, because $SO(3)$ is not simply connected, there are two homotopically inequivalent such curves, for each $R_m$ --- the corresponding doubling-up of the homotopy classes is exactly captured by the doubled-up $\Gamma^C$. Taking into account that the geometric phase corresponding to the product of two homotopy classes is the sum of the geometric phases corresponding to the factors, so that the corresponding phase factors simply multiply, we may summarize our findings in the following
\begin{theorem}
\label{thmsumm}
Consider an anticoherent state $\ket{\psi}$, the Majorana constellation of which has rotational symmetry group $\Gamma \subset SO(3)$. Then the geometric phase factors $e^{i\alpha_m}$ acquired by the rotated state $\ket{\psi(t)}=D(R(t))\ket{\psi}$, when $R(t)$ traces a path in $SO(3)$, starting at the identity $R_0$ and ending at $R_m =R_{\mathbf{n},\phi} \in \Gamma$, provide a 1-dimensional representation of the fundamental group $\pi_1(SO(3)/\Gamma)$ of the orbit space of $\ket{\psi}$, the latter group being isomorphic to the lift $\Gamma^C \!$ of $\Gamma$ in $SU(2)$.
\end{theorem}

\begin{figure}[h]
\includegraphics[width=.47\linewidth]{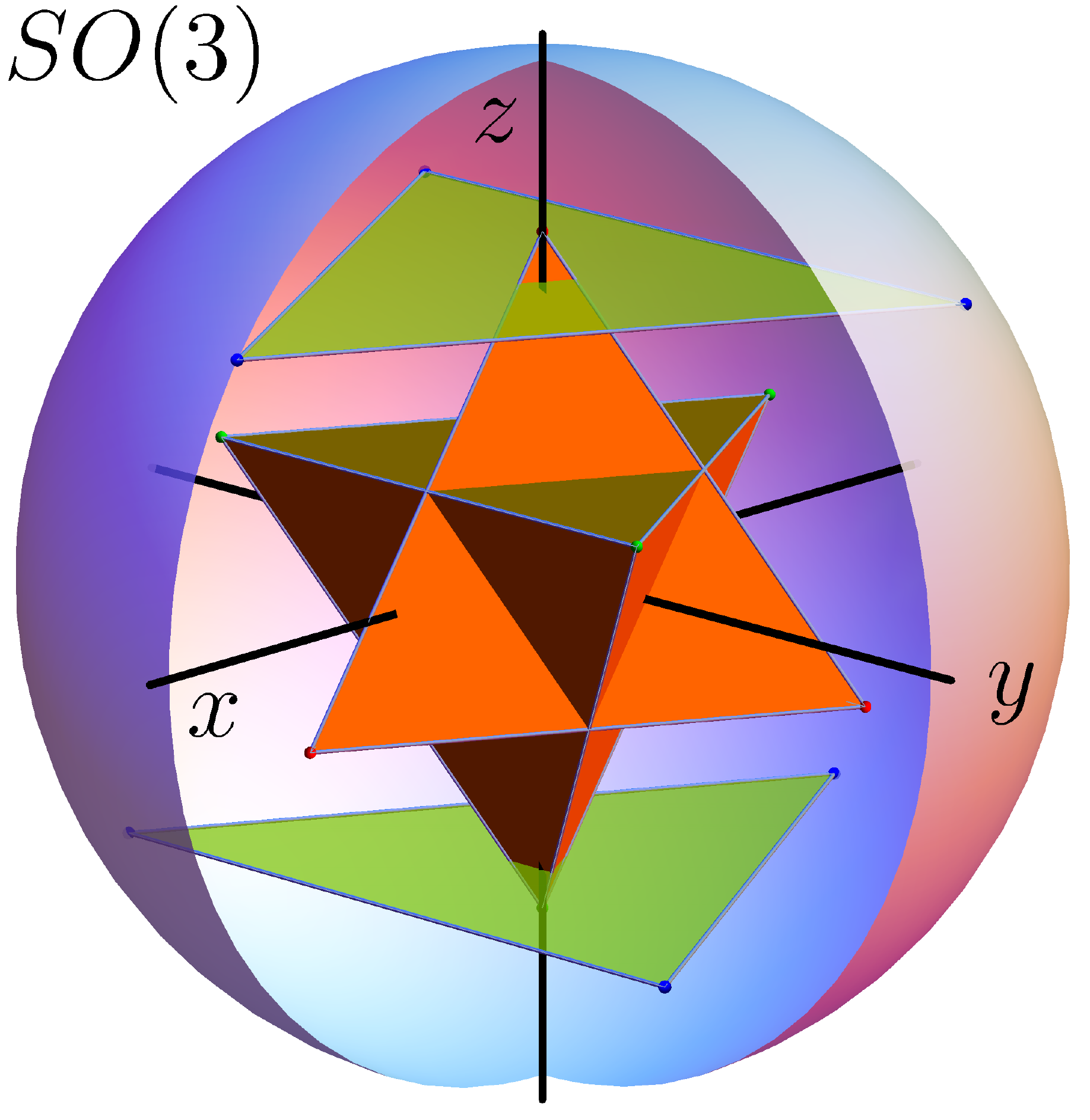}
\caption{%
Plot of the twelve elements of $\Gamma_\text{tetra}$ in the axis-angle parametrization of $SO(3)$. To help visualize their position, the twelve points have been organized as follows:  the vertices of the two tetrahedra give the four symmetry rotations by $2\pi/3$ and their inverses, while the vertices of the two triangles give the three rotations  by $\pi$ --- these latter appear twice, as pairs of antipodal points, which are identified. The identity $R_0$ is at the origin and cannot be seen in the figure. 
}
\label{fig:GammaPlot}
\end{figure}

\begin{figure}[h]
\includegraphics[width=.47\linewidth]{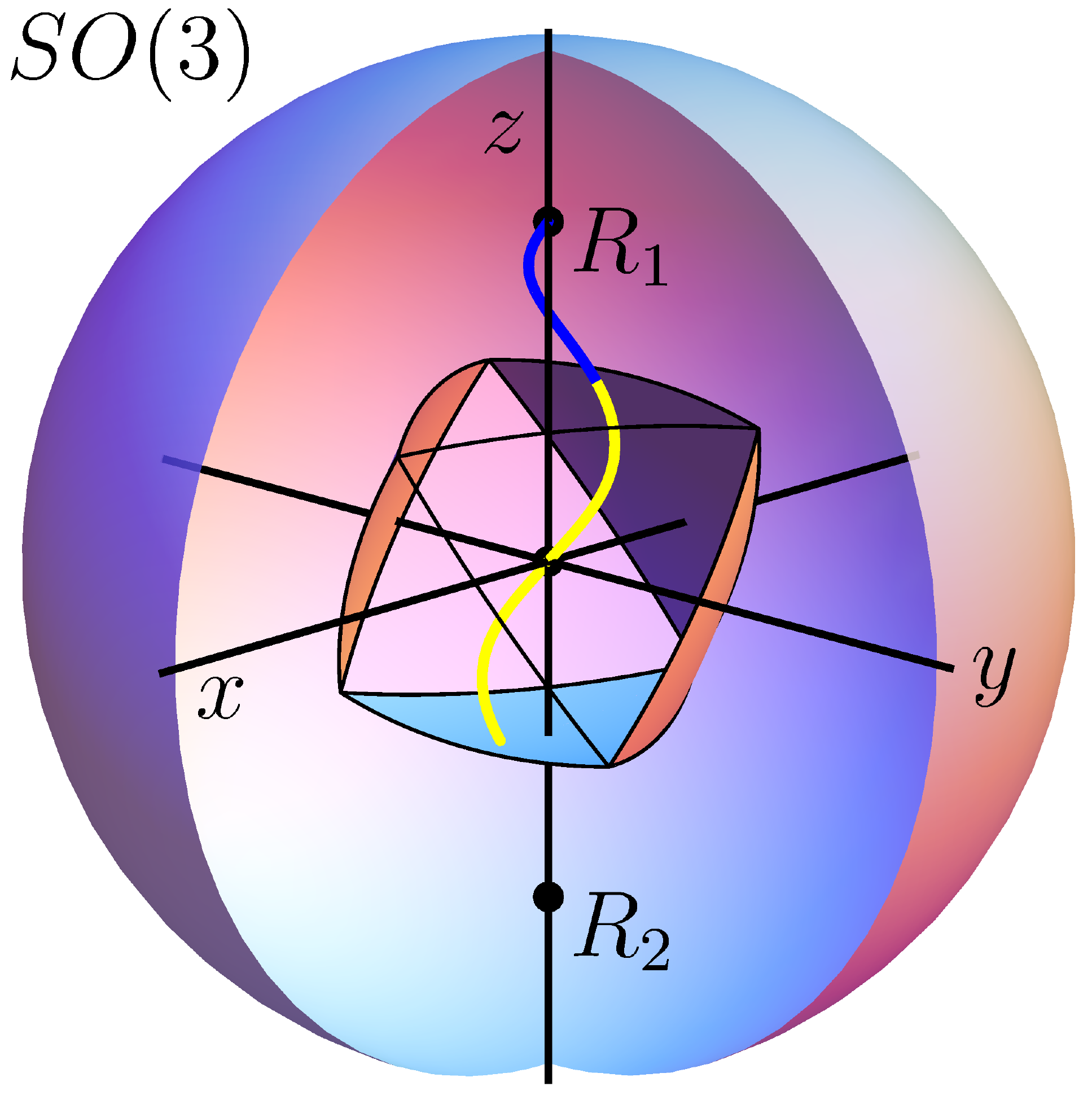}
\caption{%
The $SO(3)$ ball and the identity cell $C_0$, which may be identified with the orbit space $\mathcal{O}_{\pket{\phi_\text{tetra}}}$. Three of the faces of $C_0$ are not shown, so that its interior remains visible. A curve $C_{01}$ is plotted that starts at $R_0$ and ends at $R_1=R(\hat{z},2\pi/3)$. When the curve crosses into the cell $C_1$ (denoted by a change of color in the figure), its image $\tilde{C}_{01}$ in $\mathcal{O}_{\pket{\phi_\text{tetra}}}$ reappears at the bottom of $C_0$ and loops back at the identity, as $C_{01}$ reaches $R_1$. Thus, while $C_{01}$ is open in $SO(3)$, $\tilde{C}_{01}$ is a closed loop in $\mathcal{O}_{\pket{\phi_\text{tetra}}}$, based at the identity. The geometric phase along this loop only depends on its homotopy class, and is therefore immune to deformations.%
}
\label{fig:curves1}
\end{figure}

As a concrete example of the above general setup, consider the spin-2 anticoherent state 
$\ket{\phi_\text{tetra}}=(1,0,0,\sqrt{2},0)/\sqrt{3}$,
expressed in the $S_z$ eigenbasis $(2,1,0,-1,-2)$, with Majorana constellation given by a regular tetrahedron --- the corresponding rotation symmetry group $\Gamma_{\text{tetra}}$ contains 12 elements, shown in Fig.~\ref{fig:GammaPlot}, while the cell $C_0$, surrounding the identity $R_0$ in $SO(3)$, is shown in Fig.~\ref{fig:curves1}.
%%%%%%%%%%%%%%%%FIGURE BEGINS
%%%%%%%%%%%%%%%%FIGURE ENDS

%%%%%%%%%%%%%%%%FIGURE BEGINS
%%%%%%%%%%%%%%%%FIGURE ENDS
The geometric phases  accumulated as a result of symmetry rotations of the tetrahedral state, as well as some other interesting cases, are summarized in table~\ref{geophaseplatonic}.
%%%%%%%%%%%%%%%%  TABLE begins
\begin{table}
\begin{tabular}{| l | c | c | c | c | c |}
\hline
& 
$\phantom{\rule[-.2ex]{.5ex}{2.5ex}}$2$\phantom{\rule[-.2ex]{.5ex}{2.5ex}}$
& 
$\phantom{\rule[-.2ex]{.5ex}{2.5ex}}$3$\phantom{\rule[-.2ex]{.5ex}{2.5ex}}$ 
&
$\phantom{\rule[-.2ex]{.5ex}{2.5ex}}$4$\phantom{\rule[-.2ex]{.5ex}{2.5ex}}$ 
&
$\phantom{\rule[-.2ex]{.5ex}{2.5ex}}$5$\phantom{\rule[-.2ex]{.5ex}{2.5ex}}$ 
& 
$\phantom{\rule[-.2ex]{.5ex}{2.5ex}}$$2s$$\phantom{\rule[-.2ex]{.5ex}{2.5ex}}$
\\ \hline
spin $s$, m=0 $\rule[-.2ex]{0ex}{2.5ex}$& $s \pi$ & - & - & - & -
\\ \hline
spin $s$, GHZ $\rule[-1.1ex]{0ex}{3.4ex}$& - & - & - & - & $\pi$
\\ \hline
Tetrahedron & $0$ & $\phantom{\rule[-1.1ex]{0ex}{3.4ex}} \frac{2\pi}{3}$ & - & - & - 
\\ \hline
$\rule[-.2ex]{0ex}{2.5ex}$Cube & $0$ & $0$ & $0$ & - & -
\\ \hline
Octahedron $\rule[-.2ex]{0ex}{2.5ex}$& $\pi$ & $0$ & $\pi$ & - & -
\\ \hline
Dodecahedron $\rule[-.2ex]{0ex}{2.5ex}$& 0 & 0 & - & 0 & -
\\ \hline
Icosahedron $\rule[-.2ex]{0ex}{2.5ex}$& 0 & 0 & - & 0 & - 
\\ \hline
\end{tabular}
\caption{Absolute values of geometric phases corresponding to symmetry rotations of the spin $s$, $m=0$ and GHZ states, and those corresponding to the  platonic solids (inverse rotations give opposite phases). The rotations are specified by their order (top row), which, in this case, uniquely defines them.}
\label{geophaseplatonic}
\end{table}
%%%%%%%%%%%%%%%%  TABLE ends
These phases can, in principle,  be computed by applying the appropriate rotation matrix to the corresponding state, but, in fact, can be shown to only depend on the number of stars, in the constellation of $\ket{\psi}$, pointing in the direction of the rotation axis, and the corresponding rotation angle~\cite{Bar.Tur.Dem:07}, thus reducing their computation to simple star gazing.

Some remarks are due at this point:
\begin{enumerate}
\item
The geometric phase of $m=0$ spin-$s$ states reported in~\cite{Rob.Ber:94} is a special case of our general setting, in which the Majorana constellation consists of $s$ points in a direction $\hat{\mathbf{n}}$, and another $s$ points in the antipodal direction --- the corresponding symmetry rotation exchanges the two groups of points. Note that the symmetry group in this case has a continuous component (rotations around $\hat{\mathbf{n}}$).
\item 
The Majorana constellation of the spin-$s$ GHZ state $\ket{\psi_{\text{GHZ}}} \sim \ket{s,s}+\ket{s,-s}$ consists in $2s$ equidistant points along the equator --- the symmetry rotation is around the $z$-axis by an angle of $\pi/s$.
\item
The above phases are insensitive to perturbations of the path taken in $SO(3)$, but are still affected by end-point imprecision. However, the effect is weak, being at most quadratic in the rotation error, \ie, assuming a rotation 
\begin{equation}
\label{errorrot}
R=e^{-i\epsilon \hat{\mathbf{n}} \cdot \mathbf{S}} R_m
\end{equation} 
is applied to $\ket{\psi}$, where $R_m$ is a symmetry rotation and the prefactor is due to noise, it is easily seen that 
\begin{equation}
\label{errorphase}
\bra{\psi} R \ket{\psi} = e^{i \alpha_m}(1+\mathcal{O}(\epsilon^2))
\, ,
\end{equation}
with $\alpha_m$ as in~(\ref{Rspsi}).
\item
A further virtue of the use of anticoherent states in the setup we consider, is that the vanishing of the spin expectation value implies the absence of dynamical phase, when the states are rotated with magnetic fields --- generic states, on the contrary, do accumulate dynamical phase, which requires further processing for its elimination (see, \eg,~\cite{Fal.Faz.Pal.Sie.Ved:00,Jon.Ved.Eke.Cas:00}). Thus, if a beam of spins in an anticoherent state is split into two, and each of these secondary beams is subjected to a different symmetry rotation, say, $R_1$, $R_2$ respectively, when the two beams are brought back together to interfere, the pattern observed will only depend (apart from noise effects) on the difference $\alpha_1-\alpha_2$.
\item
Zimba gives the following generalization of the concept of anticoherence~\cite{Zim:06}: a state $\ket{\psi}$  is $k$-anticoherent if $k$ is the largest integer such that $\bra{\psi} (\hat{\mathbf{n}} \cdot \mathbf{S})^r \ket{\psi}$ is independent of $\hat{\mathbf{n}}$ for $r=1, \ldots, k$. For a $k$-anticoherent state, the error in the phase in~(\ref{errorphase}) is independent of $\hat{\mathbf{n}}$ up to $\mathcal{O}(\epsilon^k)$. For example, the tetrahedral state $\ket{\phi_{\text{tetra}}}$ turns out to be 2-anticoherent. Accordingly, if a rotation $R$ as in~(\ref{errorrot}) is applied, with $\epsilon=.1$ (about 6 degrees), the error in the phase obtained will be of the order of $10^{-2}$, with $\hat{\mathbf{n}}$-dependent part of the order of $10^{-3}$.
\end{enumerate}
We are currently working on a  generalization of the above to the non-abelian case and plan to report our findings in this direction in a forthcoming publication.
%%%%%%%%%%%%%%%%%%%%%%%%%%%%%%%%%%%%
\section*{Acknowledgements}
The authors wish to acknowledge partial financial support from DGAPA PAPIIT UNAM project 
IG100316.
%%%%%%%%%%%%%%%%%%%%%%%%%%%%%%%%%%%
%%%%%%%%%%%%%%%%%%%%%%%%%%%%%%%%%%%%%%%%%%%%%%%%%%%%%%%%%%%
%\bibliographystyle{plain}
%\bibliographystyle{unsrt}
%\bibliography{../strings}

\end{document}